\documentclass[apj]{emulateapj}
\usepackage{apjfonts}

\def\wig#1{\mathrel{\hbox{\hbox to 0pt{%
          \lower.5ex\hbox{$\sim$}\hss}\raise.4ex\hbox{$#1$}}}}

\shorttitle{Low-Mass Low-Density Transiting Planet Atmospheres}
\shortauthors{Fortney et al.}

\newcommand{\me}{$M_{\oplus}$}
\newcommand{\re}{$R_{\oplus}$}

\newcommand{\te}{$T_{\rm eff}$}

\newcommand{\teff}{$T_{\rm eff}$}
\newcommand{\teq}{$T_{\rm eq}$}

\newcommand{\tint}{$T_{\rm int}$}

\newcommand{\gj}{GJ 1214b}
\newcommand{\cp}{\citep}
\newcommand{\ct}{\citet}

\newcommand{\icarus}{Icarus} 
\newcommand{\muH}{\mu_{\rm H}}
\newcommand{\muHe}{\mu_{\rm He}}
\newcommand{\muWater}{\mu_{\rm H2O}}

\slugcomment{ApJ in press}

\begin{document}

\title{A Framework for Characterizing the Atmospheres of Low-Mass Low-Density\\Transiting Planets}

\author{Jonathan J. Fortney\altaffilmark{1}, Christoph Mordasini\altaffilmark{2}, Nadine Nettelmann\altaffilmark{1}, Eliza Kempton \altaffilmark{3}, Thomas P. Greene\altaffilmark{4}, Kevin Zahnle\altaffilmark{4} }

\altaffiltext{1}{Department of Astronomy and Astrophysics, University of California, Santa Cruz, CA 95064; jfortney@ucolick.org}
\altaffiltext{2}{Max-Planck-Institut f\"{u}r Astronomie, K\"{o}nigstuhl 17, D-69117 Heidelberg, Germany}
\altaffiltext{3}{Department of Physics, Grinnell College, Grinnell, IA, USA}
\altaffiltext{4}{Space Science and Astrobiology Division, NASA Ames Research Center, Moffett Field, CA, USA}

\begin{abstract}
We perform modeling investigations to aid in understanding the atmospheres and composition of small planets of $\sim$2-4 Earth radii, which are now known to be common in our galaxy.  GJ 1214b is a well studied example whose atmospheric transmission spectrum has been observed by many investigators.  Here we take a step back from GJ 1214b to investigate the role that planetary mass, composition, and temperature play in impacting the transmission spectra of these low-mass low-density (LMLD) planets.  Under the assumption that these planets accrete modest hydrogen-dominated atmospheres and planetesimals, we use population synthesis models to show that predicted metal enrichments of the H/He envelope are high, with metal mass fraction $Z_{\mathrm{env}}$ values commonly 0.6 to 0.9, or $\sim$100 to 400+ times solar.  The high mean molecular weight of such atmospheres ($\mu \approx 5-12$) would naturally help to flatten the transmission spectrum of most LMLD planets.  The high metal abundance would also provide significant condensible material for cloud formation.  It is known 
that the H/He abundance in Uranus and Neptune decreases with depth, and we show that atmospheric evaporation of LMLD planets could expose atmospheric layers with gradually higher  $Z_{\mathrm{env}}$.  However, values of $Z_{\mathrm{env}}$ close to solar composition can also arise, so diversity should be expected.  Photochemically produced hazes, potentially due to methane photolysis, are another possibility for obscuring transmission spectra.  Such hazes may not form above \teq\ of $\sim800-1100$ K, which is testable if such warm, otherwise low mean molecular weight atmospheres are stable against atmospheric evaporation.  We find that available transmission data are consistent with relatively high mean molecular weight atmospheres for GJ 1214b and ``warm Neptune" GJ 436b.  We examine future prospects for characterizing GJ 1214b with Hubble and the James Webb Space Telescope.  
\end{abstract}

 
\section{Introduction}
The past decade has seen a rapid increase in the number and quality of observations of transiting planet atmospheres.  Since the first observations of light transmitted \cp{Charb02} and emitted \cp{Charb05,Deming05b} by hot Jupiters, the realm of exoplanet characterization has expanded into a larger sample size of hot Jupiters \cp{Seager10} as well as into smaller planets.  Two benchmark smaller planets are the 22.4 \me\ ``warm Neptune" GJ 436b \cp{Butler04,Gillon07} and the ``super-Earth mass'' 6.5 \me\ GJ 1214b \cp{Charb09}.  Characterizing these atmospheres is a stepping stone towards one day characterizing the relatively thin atmospheres of rocky planets.

In addition, results for NASA's \emph{Kepler Mission} \cp{Borucki10,Borucki11,Batalha13} have now shown that 2-3 \re\ planets, perhaps at the boundary between super-Earths with thin outgassed atmospheres and sub-Neptunes with accreted envelopes, appear to be an extremely common type of planet \cp{Howard12,Youdin11}.  More recent work suggests that this planet size bin could be the \emph{most} common size of planet around Sun-like stars, at least on orbits less than 85 days \cp{Fressin13}, and these planets are also common around M stars \cp{Dressing13}.

\gj\ is our nearby example of a 2-3 \re\ planet, and the question of "What kind of planet is it?" is an incredibly interesting one.  It is already known that this question cannot be answered from its mass and radius alone.  Solutions that involve mixtures of rock/iron, water, and H/He gas range from models with a rock/iron core and H/He envelope, to those with massive amounts of water, no H/He, and a small rocky core \cp{Rogers10}.  \ct{Nettelmann11} and \ct{Valencia13} have suggested that water-rich models with a steam atmosphere and no H/He envelope are unlikely on cosmogonic grounds.  In models with a H/He envelope, the water (or water plus other volatiles) mixing ratio in the envelope or in the core are extremely poorly constrained \cp{Nettelmann11,Valencia13}.  As was suggested by \ct{MillerRicci09} and \ct{Kempton10}, probing the outer atmosphere of the low-mass low-density (LMLD) planets can in principle constrain the composition of the gaseous envelope.  This in turn helps to constrain the bulk composition of the planet.

Tremendous recent effort has gone into characterizing \gj\ in particular, since it is the lowest mass transiting planet with an atmosphere than can currently be characterized.  \ct{Bean10} were the first to obtain a transmission spectrum, which showed a transit radius that was essentially flat at red optical wavelengths.  Many additional observations from the ground \cp{Bean11,Carter11,Croll11,Crossfield11,deMooij12,Narita12,Teske13,Narita13} and from space \cp{Desert11,Berta12,Fraine13} have for the most part confirmed this view of a relatively featureless spectrum.

The relatively flat spectrum from the blue to the mid infrared has been suggested as being due to either a high mean molecular weight ($\mu$) atmosphere, which shrinks the scale height and transmission spectrum features \cp{MillerRicci09}, or due to some kind of gray atmospheric condensate, which would obscure molecular absorption features.  The viability of equilibrium condensates was first briefly explored in \ct{Kempton12}.  The viability of non-equilibrium condensates (such as created by the photolysis of methane and perhaps other molecules) has also been investigated.  \ct{Howe12} explored a number of photochemical hazes as a way to flatten the model spectrum.  More recently \ct{Morley13} found that a range of viable equilibrium condensates, including KCl and ZnS, and methane-derived hazes, using output from \ct{Kempton12} models, could readily flatten the planet's spectrum, even down to solar metallicity compositions.  At this point it is still fair to say that the flat transmission spectrum is degenerate with regard to its explanation. 

The discovery and atmospheric characterization of additional LMLD planets is one powerful way to break this degeneracy.  In this paper we explore the phase space of LMLD planets from $\sim3-30$ \me\ as a function of atmospheric heavy element enrichment and planetary atmospheric temperatures.  Hotter versions of GJ 1214b (which has a maximum \teq $\sim 550$ K) may be warm enough that carbon chemistry will be dominated by CO, rather than CH$_4$.  This will eliminate a pathway to haze formation and could allow molecular absorption features to be seen, if a haze is what currently obscures them in \gj.

Furthermore, with the aid of state of the art population synthesis planet formation models \cp{Mordasini12a,Mordasini12b} we also investigate the expected abundance of metals (or ``heavy elements") in the atmospheres of these LMLD planets, and explore how a wide range in the mass fraction of metals in the envelope and atmosphere ($Z_{\rm env}$) will lead to changes in the transmission spectrum, as a function of planet mass.

We show that a high $Z_{\rm env}$, and corresponding high atmospheric mean molecular weight ($\mu$) appears to be a natural outcome of the planet formation process of small planets.  As planet masses increase beyond that of GJ 436b, the predicted metal enrichment of the atmospheres decreases, thereby decreasing the mean molecular weight, which would also allow molecular absorption features to be better seen.

\begin{figure} \epsscale{1.22}
\plotone{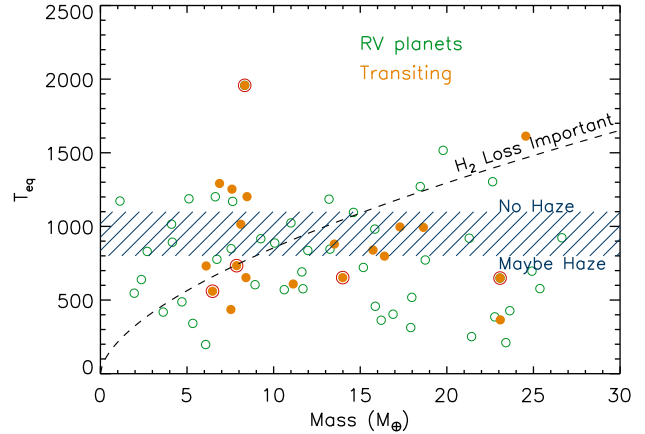}
\caption{Planetary equilibrium temperature (\teq, assuming zero Bond albedo and planet-wide average conditions) vs.~mass for radial velocity and transiting planets.  ($M \sin i$ is plotted for the radial velocity planets.)  Transiting planets with a red circle are found around bright stars, from left to right:  GJ 1214b, HD 97658b, 55 Cnc e (near top), GJ 3470b, and GJ 436b.  The hashed region marks a boundary between planets that may have hazes created by the destruction of methane, and those that do not.  (See \S \ref{haze}.) The \teq\ values are calculated by assuming planet-wide re-radiation of absorbed flux and zero Bond albedo.  The dashed black curve is from \ct{Lopez13}.   This is a relation for planets that are 5\% H/He envelope by mass, a typical LMLD planetary value.  Warmer planets above this curve may lose more than half of their envelope mass to evaporative mass loss.
\label{tempmass}}
\end{figure}

The goal of the paper is to provide a pathway for current and future planetary characterization over a wide phase space of planetary temperature, mass, and envelope composition.  While previous works, such as \ct{Spiegel10} have investigated the structure and spectra of atmospheres in the Neptune mass range, here we are interested in a wider mass range, with a strong focus on composition and diversity.  As the sample size of characterized low-mass low-density planets grows we will be able to determine the roles (if any) of atmospheric clouds/hazes and high $Z_{\rm env}$ in making transmission spectra of LMLD planets challenging.  This is a diverse population.  Figure \ref{tempmass} shows the range of planet masses vs.~planet equilibrium temperatures for transiting and radial velocity planets below 30 \me.
\section{Atmosphere Models}
\subsection{Temperature Structure and Transmission Spectra}
We have computed models of the atmosphere of \gj\ and similar planets using a collection of model atmosphere methods that have been widely used to model the atmospheres of exoplanets, brown dwarfs, and solar system planets.  The fully non-gray atmosphere code has been used to derive model pressure-temperature (\emph{P--T}) profiles for the atmosphere of Titan \cp{Mckay89}, Uranus \cp{MM99}, brown dwarfs \cp{Marley96,Burrows97,Saumon08}, and irradiated exoplanets \cp{Fortney05,Fortney08a,Cahoy10,Morley13}.  The radiative transfer methods are described in \ct{Toon89}.  We use the opacity database of \ct{Freedman08}, the ``correlated-k'' method of opacity tabulation described in \ct{Goody89}, the equilibrium chemical abundances tables of K.~Lodders and collaborators \cp{Lodders02,Lodders06,Lodders09}, and the solar system abundances of \ct{Lodders03}.

\begin{figure} \epsscale{1.22}
\plotone{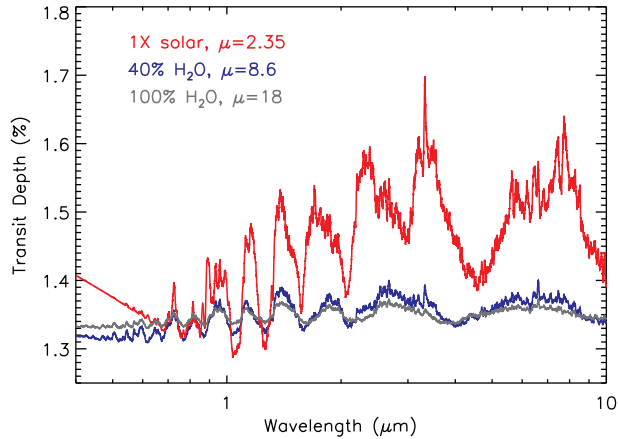}
\caption{Several transmission spectrum models for \gj.  As is well known, the dramatically larger variation in transit depth for the 1$\times$ solar models is predominately due to the low mean molecular weight ($\mu$) and corresponding larger scale height for these atmospheres.  Models in blue and gray mix in water vapor, 40\% by number, and 100\% by number, leading to $\mu$ increases from 2.35 (solar mix) to 8.6 and 18, respectively.
\label{full}}
\end{figure}

The code we use to model transmission spectra is fully described in \ct{Hubbard01}, \ct{Fortney10b}, and \ct{Shabram11}.  In \ct{Shabram11} the code was tested against the analytical relations of \ct{Lecavelier08a}.  The transmission spectrum code uses wavelength-by-wavelength opacities from the \ct{Freedman08} database.  To showcase baseline models, we have computed transmission spectra using the planet-wide average one-dimensional solar metallicity \emph{P--T} profile from \ct{Kempton10}.  We use this model, and also a version of it where the mixing ratio of H$_2$O is increased to 0.4, and a version where the water mixing ratio is 1.0 (pure steam).  These models are shown in Figure \ref{full}.

As already noted by several authors, the \gj\ model with a solar metallicity cloud-free atmosphere is ruled out.  \ct{Berta12}, fitting to the highest S/N ground-based and space-based observations, suggested that water mixing ratio greater than 0.4 best matched their \emph{HST} spectrum and all other available data, and we plot such a model, along with a pure water (steam) model for comparison.  Additional observations since \ct{Berta12} have not materially changed their constraints.  The plot shows the effect of increasing values of $\mu$, which we will discuss later in the paper.
 
 \subsection{Role of Photochemical Hazes?} \label{haze}
We wish to examine at what planetary temperatures (and hence, at what level of incident flux) pathways towards haze formation may be most likely.  \ct{Zahnle09b} and \ct{Kempton12} have shown that the photochemical destruction of CH$_4$ in atmospheres just cooler than the hot Jupiters (\te $\sim1000$ K or less) can lead to the creation of higher order hydrocarbons, which may progress to the eventual formation of ``soot'' or ``haze'' particles that could obscure the atmosphere \cp{Howe12,Morley13}.  Such haze is seen in Jupiter's atmosphere, particularly at the poles \cp[e.g.,][]{Rages99}.

\begin{figure}  \epsscale{1.22}
\plotone{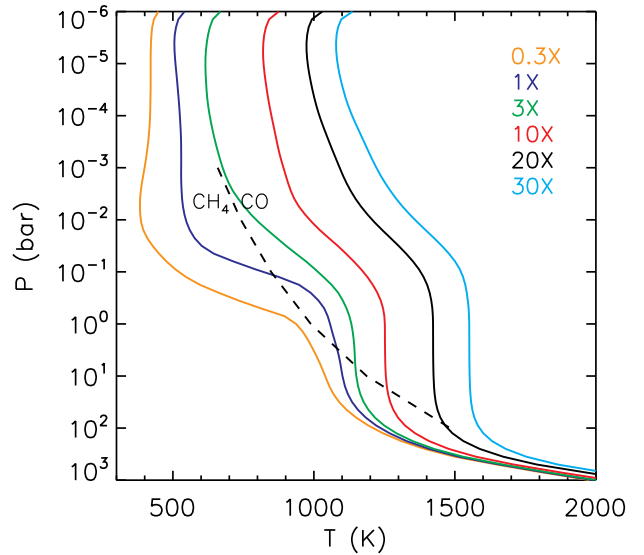}
\caption{One dimensional pressure-temperature profiles for models of GJ 1214b analogs.  Models receive levels of incident flux that vary from 0.3 to 30$\times$ that received by the actual planet.  All use 50$\times$ solar metallicity.  Planet-wide redistribution of energy is assumed, and \tint=75 K.  The dashed curve shows where CH$_4$ and CO have equal abundance in equilibrium chemistry, assuming 50$\times$ solar metallicity.  The \teff\ of the models are 412 K, 557 K, 733 K, 991 K, 1178 K, and 1303 K. 
\label{pt}}
\end{figure}

We have calculated \emph{P--T} profiles for mature \gj-like planets at a range of incident flux levels, from 0.3$\times$ to 30$\times$ that currently experienced by \gj.  These profiles utilize a metallicity that is 50$\times$ solar and the GJ 1214A parent star.  These models are shown in \mbox{Figure \ref{pt}}, compared to the chemical equilibrium equal abundance curve of CH$_4$ and CO.  Atmospheres that are everywhere to the right of this curve would be expected to be CO-dominated, with a relatively small CH$_4$ mixing ratio.
\begin{figure}  \epsscale{1.22}
\plotone{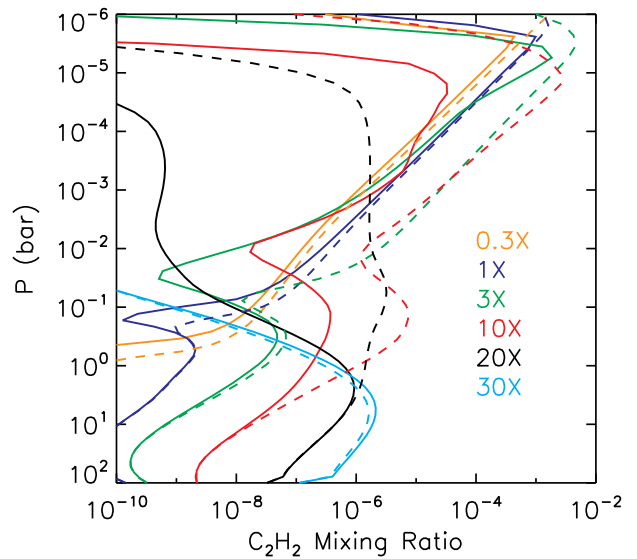}
\caption{Mixing ratio of C$_2$H$_2$ (acetylene) for the \emph{P--T} profiles in Figure \ref{pt}.  As in \ct{Morley13} the GJ 1214A stellar spectrum is used for the incident UV flux.  Calculations with two values of $K_{\rm zz}$, the eddy diffusion coefficient are plotted.  Solid curves are $10^8$ cm$^2$ s$^{-1}$ while dashed curves use enhanced mixing, $10^{10}$ cm$^2$ s$^{-1}$.  Near 10$^{-5}$ bars, the dashed profile, at an incident flux 20$\times$ higher than \gj (black), shows  a dramatic decrease in acetylene.  A similar decrease for the solid profiles is seen in the red model at 10$\times$ higher flux.  At these low pressures we find a change in C$_2$H$_2$-rich to -poor atmospheres at \teq\ between $\sim800-1100$ K.
\label{c2h2}}
\end{figure}

We extend the non-equilibrium chemistry calculations of \ct{Kempton12} and \ct{Morley13} here for versions of \gj\ at higher and lower incident fluxes.  The models were run using 50$\times$ solar metallicity and two values of the eddy diffusion coefficient, $K_{\rm zz}$.  Our aim is to understand better at what incident flux level the atmospheric carbon chemistry changes from being CH$_4$-dominated to CO-dominated.  As described in \ct{Zahnle09b} and \ct{Morley13}, the photolysis of CH$_4$ leads to the creation of C$_X$H$_Y$, higher order hydrocarbons, which are precursors to the creation of upper atmosphere aerosols, which may be described as ``hazes'' or ``soots."  Since it has been demonstrated that such hazes can blanket the transmission spectrum \cp{Howe12,Morley13} it is important to understand the planetary temperature beyond which this regime is passed.

Figure \ref{c2h2} shows the mixing ratio of acetylene (C$_2$H$_2$), the simplest higher order hydrocarbon.  We can examine its abundance as a function of \teq\ as a proxy for the abundance of all higher order hydrocarbons.  These calculations use two values of the eddy diffusion coefficient, $K_{\rm zz}$, which parameterizes the vigor of vertical mixing.  Higher values of $K_{\rm zz}$ here lead to higher mixing ratios of CH$_4$ transported from depth to the upper atmosphere, which can then be destroyed by UV flux.  Therefore any boundary in \teq\ between potentially haze-poor and haze-rich models depends on the unknown $K_{\rm zz}$ value.

For the high $K_{\rm zz}=10^{10}$ cm$^2$ s$^{-1}$ case, the warm versions of \gj\ show abundant  acetylene up until the model with 10$\times$ higher incident flux, beyond that boundary the acetylene abundance in the upper atmosphere drops off precipitously.  For the lower $K_{\rm zz}=10^{8}$ cm$^2$s$^{-1}$ case, the boundary is at lower incident fluxes, between the 3$\times$ and 10$\times$ flux models.  Taken together, these models suggests a boundary in \teq\ at $\sim800-1100$ K where the potential effects of a haze layer could wane.  This echoes the results of \ct{Zahnle09b}, who suggested a boundary of 1200 K, based on isothermal \emph{P--T} profiles. We therefore suggest that the hottest LMLD planets may not posses haze layers that could blanket their transmission spectrum, although further work on these complex photochemical problems is needed.  In particular, extremely metal-rich atmospheres may be CH$_4$ poor even at low temperatures below 800 K \cp{Moses12} .  Also a wider variety of parent stars could be considered, although planets around red dwarfs are the most amenable to characterization.

\section{Formation and Composition of LMLD Planets}
The planet formation process is chaotic and complex, as planetesimals merge to become embryos, and then planets.  The composition of the planets themselves can change as smaller objects are accreted before the gas and planetesimal disk is cleared away.  Our understanding of the accretion of planets is still developing, but we are at the point now where it is possible to directly compare models of planet formation to the distribution and composition of detected planets.  What we are interested in is understanding what $Z_{\rm env}$ may be for GJ 1214b, how this value may vary for other planets of similar mass, and how $Z_{\rm env}$ will vary as a function of planet mass. 
\begin{figure*}  \epsscale{1.1}
\plotone{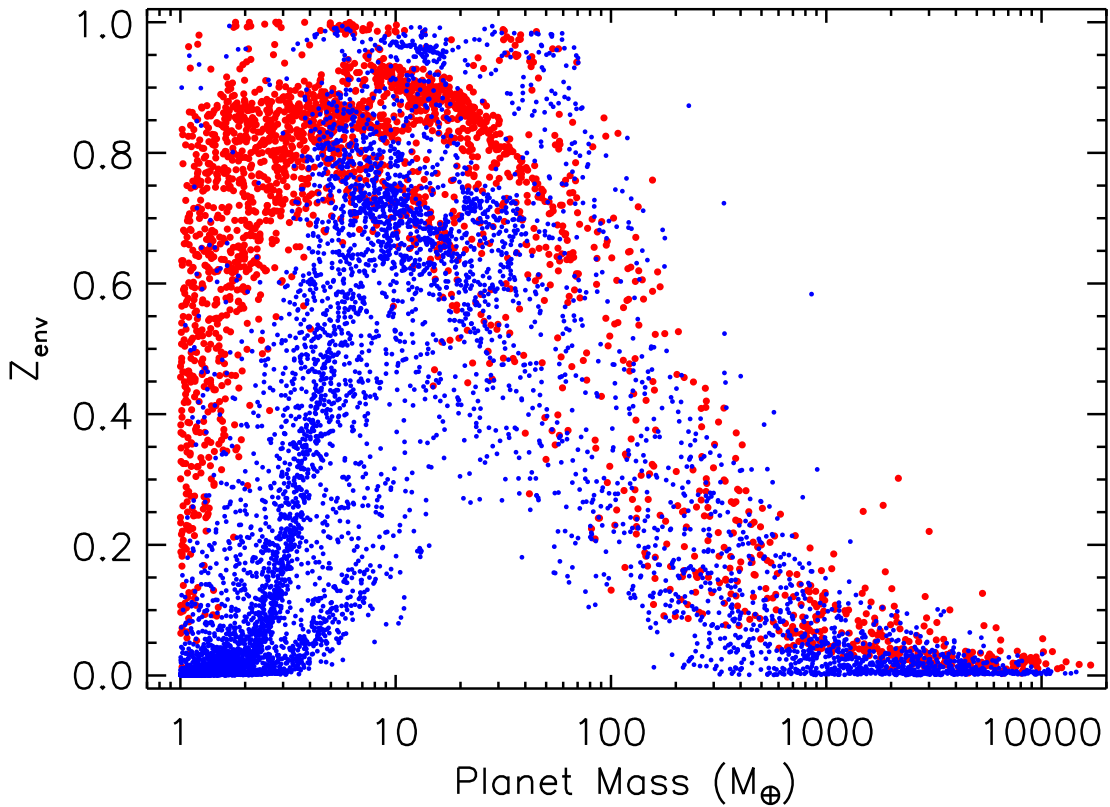}
\caption{Heavy element mass in the H/He envelope ($Z_{\rm env}$) as a function of planet mass for the output of the population synthesis models.  Blue dots use 100 km planetesimals and red dots use 1 km planetesimals.  We make a simple assumption of a uniform $Z_{\rm env}$ throughout the envelope.
\label{zenv}}
\end{figure*}

\subsection{Population Synthesis Models:  Background and Implementation}
We have conducted population synthesis calculations to obtain theoretical predictions for the heavy element fraction in H/He envelopes. The population synthesis calculations are very similar to the ones presented in \ct{Mordasini12b}. The planets are synthesized using an upgraded version \cp{Mordasini12a} of the planet formation model originally presented in \ct{Alibert04}. The model combines several simple, standard prescriptions for the growth of the solid core and the gaseous envelope \cp{Pollack96}, the evolution of the protoplanetary disk \cp{Lynden74,Papaloizou99}, and disk migration \cp{Lin86,Paardekooper10}. The reader is referred to \ct{Alibert05} and \ct{Mordasini12a} for details.  Here in particular we focus on the accretion of solid planetesimals by the protoplanetary atmosphere.

During each calculation of the formation of one synthetic planet, we simulate at every numerical timestep a collision between the protoplanet and a planetesimal of a given size. This is done using the core mass and envelope structure of the protoplanet at this moment in time. The collision is simulated with a semi-analytical model with a basic structure that is inspired by the work of \ct{Podolak88}, but takes into account the results from studying the impact of SL9 with Jupiter. This event illustrated both the importance of mechanical disruption for massive impactors \cp[e.g.,][]{Zahnle94, Boslough94}, and the lower efficiency of thermal ablation for such large bodies \cp{Field95,Svetsov95}. A short overview of the model is given in \ct{Alibert05}, \ct{Mordasini06}, and below. A more detailed description will be given in future work.  

To obtain the distribution of where mass and energy is deposited into the envelope, we integrate the 3D trajectory of the impacting planetesimal under the actions of gravity (in the two body approximation), gas drag, thermal ablation, and mechanical disruption (flattening and fragmentation). The planetesimal is described by its position, velocity, mass, and radius.  These quantities are evolved with a coupled set of differential equations. For the thermal ablation, we have developed a prescription for the heat transfer coefficient as a function of the flow regime. We solve the normal shock wave jump conditions as in \ct{Chevalier94} to get the shock wave radiation. This controls the efficiency of thermal ablation during hypersonic flight in the denser parts of the atmosphere, which is the decisive regime for large impactors \cp{Opik58,Zahnle92}.

For the mechanical disruption, we use the "pancake" model \cp{Zahnle92}, but couple it to a model for the growth of Rayleigh-Taylor instabilities that develop on the front side of the impactor that behaves approximately as a fluid at dynamic pressures that exceed significantly the impactor's tensile strength or self-gravity \cp[e.g.,][]{Roulston97,Korycansky00} . The model was tested by comparing it to simulations of different impacts, namely of the Lost City meteorite \cp{ReVelle79}, the Tunguska event \cp{Chyba93},  impacts of km sized impactors into Venus \cp{Zahnle92}, and most importantly, the SL9 collision with Jupiter \cp[e.g.,][]{Zahnle94, Boslough94}.  
 
The calculation of the trajectories of the planetesimal yields the radial mass deposition profile. We thus know the fraction of the accreted solids that is deposited in the envelope while the rest reaches the core directly. This is done during the entire formation of the planet, giving at the end of a simulation the final core mass, the total mass deposited in the envelope, and the amount of gas H/He that was accreted. 

Two populations of synthetic planets were calculated which only differ in the assumed size of the planetesimals. The size of the planetesimals is fixed during one synthesis and always equal to either 1 or 100 km. This is clearly an idealization, since in reality a size spectrum of different planetesimals will be accreted. The size spectrum will be a function of time and the distance from the star. If a protoplanet accretes beyond the iceline, the planetesimals are assumed to have an icy composition, otherwise they are made of rocks. This influences their density, tensile strength, heat of vaporization, and some other material parameters \cp{Opik58, Podolak88, Boslough94}. The impact geometry is head-on. The initial velocity of the planetesimals relative to the planet is equal to the  local escape speed, i.e., we assume that the random velocities are small.

Diversity in the formation conditions of the planets, which are manifest in the final envelope enrichment of the planets, comes from a variety of factors, as described in \ct{Mordasini12a}.  These include diversity in: the dust to gas ratio in the protoplanetary disk, according to observed [Fe/H] distributions in the solar neighborhood, the protoplanetary disk gas mass, according to observed disk mass distributions in rho Ophiuchi, the external photoevaporation rate which sets together with viscous evolution the disk lifetimes, chosen in a way that the disks have a lifetime distribution similar as observed, and finally the initial position of the embryo, uniform in the logarithm of the semimajor axis.

For the resulting envelope enrichment we make a number of assumptions (besides the general model assumptions, like no mass loss or accretion by the planets after the protoplanetary gas disk is gone, see Mordasini et al. 2012a). First we assume that the matter that is ablated during the passage of the planetesimal does not later on sink to the core. The actual fate of the solids will depend on the relative timescales of settling and mixing, which are controlled by the solubility of heavy elements in the H/He \cp{Wilson12} and the state of the envelope (radiative vs. convective layers). A sinking of a part of the solids initially deposited in the envelope would reduce the envelope metallicity.  Alternatively it is also possible that the solids initially (during the formation phase) settle deep into the envelope, but at a later stage get mixed homogeneously through the envelope \cp{Iaroslavitz07}.  Second, we assume that the solid core (formed by impactor material that can directly penetrate through the entire envelope) does not subsequently dissolve \cp[see][]{Guillot04,Wilson12}. This process would in contrast increase the envelope metallicity.

Finally, we caution that the calculations are currently not self-consistent since our planetary structure code is not yet able to handle a compositionally varying equation of state and opacity. While we keep track of the amount of solids deposited in the envelope, we nevertheless add all solids to the (computational) core mass. This set-up corresponds to a similar simplification made in \ct{Pollack96}.  Thus the formation and evolution of the planets is in fact the one of planets with all solids in the (computational) core and a pure H/He envelope which is modeled with the EOS of \ct{SCVH}, and the solar composition opacities of \ct{Bell97} and \ct{Freedman08}. This must be critically kept in mind, since it has been shown \cp{Hori11} that the enrichment of the envelope  lowers the critical core masses and speeds up gas accretion \cp[see also][]{Stevenson82b}. This means that the envelope composition feeds back onto the formation process.

One should further note that compositional gradients can potentially have important implications also for the evolution of planets by inhibiting efficient, large-scale convection \cp{Stevenson85, Leconte12}. We finally mention that the formation model, including the part used to simulate the impact, are strongly simplified descriptions of the actual processes. This means that the resulting envelope enrichments must be regarded as first, rough estimates. In particular the link between the initial deposition in the envelope, and the atmospheric enrichment after several Gyrs of evolution needs to be explored.

\subsection{Population Synthesis Models:  Outputs}
In Figure \ref{zenv}, the blue dots show the envelope enrichment using 100 km planetesimals as a function of planet mass, while the red dots assume 1 km planetesimals.  For 100 km planetesimals, and  the lowest planetary masses (less than a few Earth masses),  $Z_{\rm env}$  is typically relatively low. This is a consequence of the following: These very low mass planets can only accrete tenuous H/He envelopes since the gas accretion timescale nonlinearly increases for lower core mass \cp[e.g.,][]{Ikoma00}. The 100 km planetesimals can fly through these thin atmospheres without losing much mass, but instead deposit their entire mass directly in the core, leading to low enrichments. While this basic mechanism is plausible (cf. the radar-dark ``shadows'' on Venus indicating that the planet is ``opaque" to planetesimals smaller than a certain critical size, Zahnle 1992), this is not necessarily a realistic result: If there are additionally smaller planetesimals which contribute important amounts of solids, they would get destroyed already in the tenuous envelopes.  This is indeed the case for the 1 km planetesimals. In the simulations with these smaller impactors, the very low mass planets are found to have high primordial enrichments. 

For the 100 km planetesimals, $Z_{\rm env}$ then grows with mass to a local maximum of approximately 0.7 at a planet mass of  $\sim$7 \me.  This is due to the fact that these planets can accrete a sufficient amount of gas so that the planetesimals get efficiently destroyed in the envelope, but  their gas envelope is still not very massive, leading to the high values. For even more massive planets, $Z_{\rm env}$ starts to decrease again, since now the mass of H/He that can be accreted increases. Giant planets have low $Z_{\rm env}$, as expected, because gas runaway accretion strongly dilutes the metals in the envelope. The results for the 1 km planetesimals are qualitatively similar, with the difference that the $Z_{\rm env}$ are usually higher, as expected. An important common feature of the two populations is that at all masses, there is a wide spread in possible $Z_{\rm env}$. If this can be preserved and has an imprint on the atmospheric composition we can observe today, we expect to find a large compositional diversity, too.  This diversity would overlay the general trend of a decreasing envelope enrichment with planet mass characterizing \mbox{Figure \ref{zenv}}.

\subsection{Envelope Metallicity}
A key question that one can examine from the population synthesis output is how $Z_{\rm env}$ changes with planet mass.  Clearly Figure \ref{zenv} shows a tremendous range of planetary properties.  This is a clear example of the planetary diversity that should at a minimum be expected.

\begin{figure}  \epsscale{1.22}
\plotone{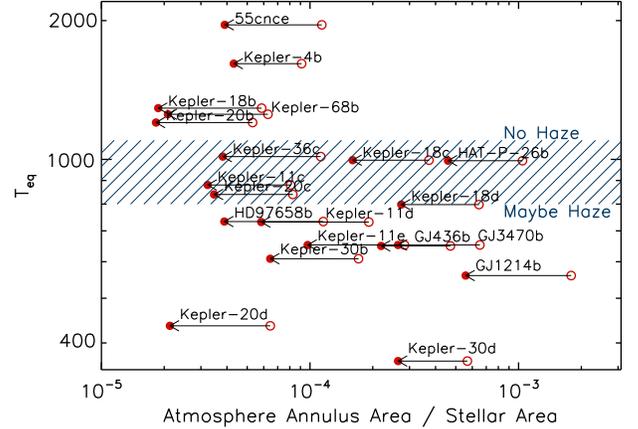}
\caption{Planetary equilibrium temperature plotted as a function of the atmospheric annulus area ratio.  This ratio is defined as the cross-sectional area of five scale heights of atmosphere divided by the cross sectional area of the parent star.  Open circles use a temperature of \teq\ and a mean molecular weight $\mu=2.35$, appropriate for a solar composition gas.  Fill circles instead use a larger $\mu$ taken from the planet formation simulations.  The larger $\mu$ leads to a more compact atmosphere.
\label{temparea}}
\end{figure}

The quantity of interest when trying to determine the suitability of a planet for characterization by transmission spectroscopy is a comparison of the area of the atmosphere's annulus to the cross-sectional area of the parent star.  In Figure \ref{temparea} we examine this area ratio for all known transiting low mass planets.   We have arbitrary used an atmospheric temperature equal to \teq\ and a height of five scale heights to plot the ratio of the area of the atmosphere annulus to the stellar area.  Open circles use the $\mu$ of a solar mix of 2.35, while filled circles use a simply estimated higher $\mu$, as a function of mass, based on Figure \ref{zenv}.   We can crudely estimate a mean $Z_{env}$ as a function of planet mass, further assume it is uniform in the entire envelope, and then convert $Z_{env}$ to $\mu$.  We do this by assuming that the heavy element component is water, to find the average $\mu$ as a function of planet mass.   The value of $\mu$ for some planets reaches 7, dramatically shrinking the scale height.   The arrows show the shift for individual planets, which are labeled.  The ratio for \gj\ falls by a factor of $\sim$~3.5, although we note that it is still the most promising target.  We note that the apparent magnitude of the parent star is also a particularly important factor in the suitability of characterizing particular planets.

\begin{figure*}  \epsscale{1.1}
\plotone{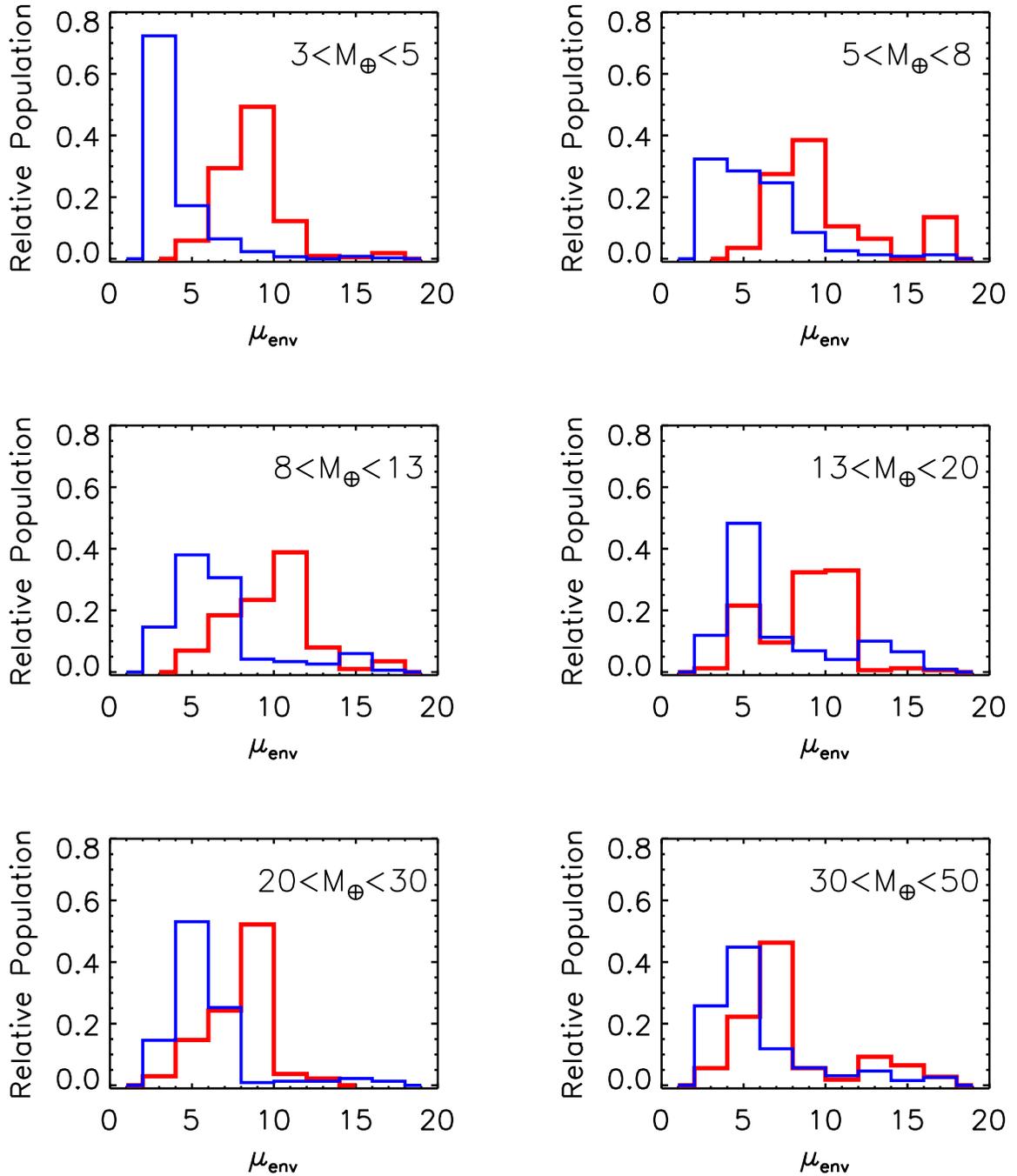}
\caption{Histogram of the mean molecular weight ($\mu$) of the H/He envelope of planets formed in the population synthesis models.  Heavy elements were assumed to be pure water.  Blue lines use 100 km planetesimals and red lines use 1 km planetesimals.  Groups of planets are broken up by mass.  There is a weak trend towards lower $\mu$ at higher planet masses in this sample.  For the group of planets with masses similar to GJ 1214b (5 to 8 \me) values of $\mu$ of $\sim$8 are quite common.
\label{histo}}
\end{figure*}

Of course Figure \ref{temparea} is just one realization of the wide range of possible values for $\mu$ for these planets.  We have taken the population synthesis output of Figure \ref{zenv} and binned the output to find the range of $\mu$ values for different planetary mass bins.  Figure \ref{histo} shows our results.  Several trends are worth noting.  For 100 km planetesimals (in blue) the lowest-mass bin (3 to 5 \me, upper left) values of $\mu$ are close to solar, since these planetesimals readily pass through the envelope without ablating.  Only for the larger planetary mass bins does $\mu$ begin to increase, due to the greater likelihood of appreciable deposition of the heavy elements into the H/He envelope.  For the 1 km planetesimals (in red) values of $\mu$ of $\sim$10 are common from 3 to 20 \me, with a modest trend to lower $\mu$ for the highest mass bin.  In nearly all cases and for nearly all envelopes, the 1 km planetesimals dissolve into the H/He envelopes, enriching these envelopes.

One should clearly expect a tremendous amount of diversity in $\mu$ for LMLD planets, in particular below 15 \me.  Put another way, at this point there is no reason to expect that all planets with the mass of \gj\ should show a similar atmospheric metallicity enrichment.  The size of accreted planetesimals, and their composition, could vary greatly based on planetary \teq, stellar type, formation time, the mass and location of nearby planets, and other parameters.

\subsection{Comparison with Uranus and Neptune} \label{unsec}
Within the solar system the planets Uranus and Neptune are the closest to LMLD planets in log(mass).  While it could be an overstatement to suggest that LMLD planets are all smaller version of these planets, it is worthwhile to review our current understanding of the structure and atmospheres of these planets.  A window into interior structure that is only available in the solar system is the gravity field of the planets, which yields constraints on planetary density as a function of radius.  Recently, \ct{Nettelmann13} have investigated constraints on the interiors structure, including uncertainties in the gravity fields, shapes, and rotation rates of the planets.  ``Standard'' three-layer models were calculated, including a rock/metal core, a middle layer composed mostly of water, and an outer layer.   The water mass fraction of the outer envelope ($Z_\mathrm{env}$) of up to 0.7 are acceptable for Neptune, and 0.2 for Uranus.
\begin{figure} \epsscale{1.1}
\plotone{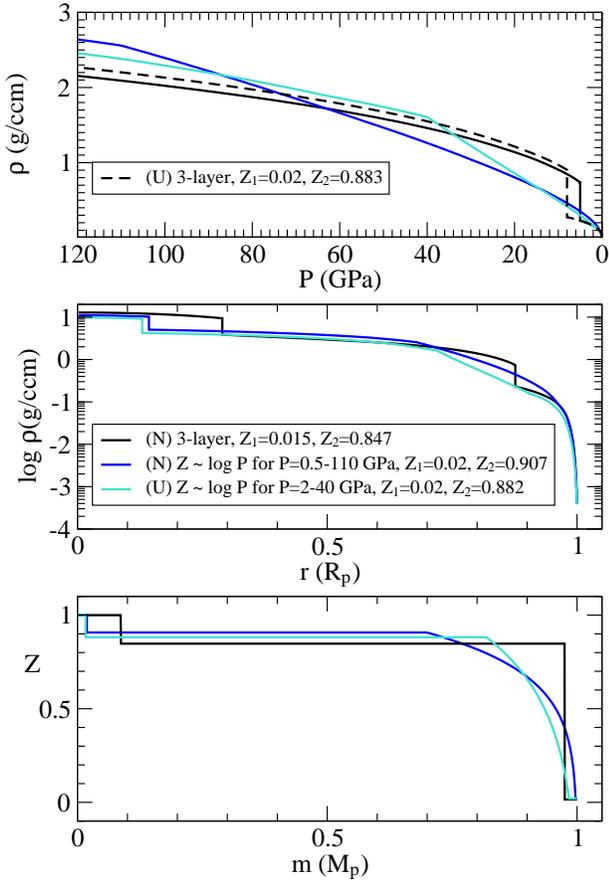}
\caption{Non-standard interior models of Uranus and Neptune that match all observational constraints.  The top panel shows density vs.~pressure in the outer envelope, the middle panel density vs.~radius for the entire planet, and the bottom panel $Z$ vs.~mass shell for the entire planet. While standard models (black, for Neptune) include a small water-poor outer layer and sharp transition to water-rich layer, one can also derive models for the planets (light blue for Uranus and dark blue for Neptune) that feature a smooth transition to larger $Z$ values with depth.
\label{UN2}}
\end{figure}
It is tempting to compare these numbers directly to the model productions of $Z_{\rm env}$ from the population synthesis models.  However, this must be done with great care.  Such three layer models are likely artificial, and if the interior structure is more complex (see \S \ref{evap}) it is no longer clear what $Z_{\rm env}$ for these planets should be.

One can also compare the molecular abundances in the atmosphere of Uranus and Neptune to the elemental abundances in the Sun.  Since the planets are quite cold, most volatile elements are condensed into clouds far below the visible atmosphere.   Methane is only partly condensed, so it is possible to derive its abundance in the troposphere from infrared spectroscopy.  Data from \ct{Baines95}, along with the revised solar abundances \cp[e.g.,][]{Lodders03}, suggest an enrichment of carbon of $54 \pm 15 \times$ solar for Neptune \cp{Fletcher10} and similar number of $\sim50 \times$ solar for Uranus.  Furthermore, the detection of CO in the atmosphere of Neptune, brought up from deeper layers by vertical mixing, suggests an even larger mixing ratio of water.  Past and current modeling work \cp{Fegley94b,Luszcz13} suggest that the atmospheric oxygen abundance of Neptune could be as large as $\sim400-600\times$ solar.  Assuming a $\mu$ for the heavy elements of 18 amu (correct for water) $55\times$ solar would lead to $Z=0.40$ and $500\times$ solar is $Z=0.95$ \cp[see, for instance Figure 6 of][and our Appendix]{Nettelmann11}.  These very high values of $Z$ in the H/He atmosphere of Neptune are similar to what was found from the gravity fields of the planet \cp[$Z$ values up to 0.7 are allowed][]{Nettelmann13}.  This suggests thats it is reasonable for GJ 1214b and LMLD planets as a class to have a similar $Z$ in their H/He envelopes, or perhaps even higher.

\section{Atmospheric Escape}
There are several lines of observational evidence that H$_2$ dominated exoplanet atmospheres are not stable against atmospheric evaporation.  Most prominent are the observations that hot Jupiters lose atmospheric mass \cp{Vidal03,Linsky10,Lecavelier10,Lecavelier12}, as has been observed in the ultra-violet with the \emph{Hubble Space Telescope}.  A more subtle but emerging trend is the incident flux vs.~planet density plane, which shows a lack of LMLD planets at high incident fluxes \cp{Lopez12,Lopez13}.  At the time of the discovery of \gj, it was clear that atmospheric evaporation could be important in shaping the planet that we see today \cp{Charb09,Rogers10,Nettelmann11}.

\subsection{Hydrogen Loss Carries Heavier Elements}
Since it is likely that LMLD planets lose mass, and that a high $\mu$ atmosphere could explain the \gj\ transmission observations, it is natural to wonder if \emph{differential} mass loss can occur.  If hydrogen is lost preferentially, the atmosphere could become enriched in heavy elements with time.   However, we find that this is not possible for a hydrogen-dominated atmosphere. \cp[See, e.g.][for a review.]{Hunten73}.  We show this by examining the diffusion-limited flux for hydrogen with respect to other gases. The maximum escape flux of H$_2$ (component $a$, for atmosphere) through a non-escaping minor gas (component $i$) is: 
\begin{equation}
n_a u = b_{ia} \frac{(m_i-m_a)GM}{kTR^2}
\end{equation}
 \begin{equation}
\label{eight}
n_a u = \phi_{lim} = {\left(m_i-m_a\right)gb_{ia}\over kT}{f_a} - {\alpha_{ia}b_{ia}\over T}{dT\over dz}{f_a}.
\end{equation}
where $n_a u$ is the H$_2$ flux , $b_{ia}$ (cm$^{-1}$s$^{-1}$) is the binary diffusion coefficient between H$_2$ and the minor gas,
 and $m_a$ and $m_i$ are the masses of H$_2$ and the minor gas, respectively. For H$_2$ and either CO or CH$_4$, typical 
minor gases, the binary diffusion coefficient at relevant temperatures is $b_{ia}=4.6\times10^{19}(T/1000)^{0.75}$cm$^{-1}$ \cp{Marrero72}.  The time scale for diffusion-limited H$_2$ escape of a giant planet is $\tau_{dl}$ = $M/ \dot M$. 
 With $\dot M = 4 \pi m_a n_a u R^2$, $\tau_{dl}$ becomes:
\begin{equation}
\frac{M}{\dot M} = \frac{kT}{4 \pi m_a b_{ia} (m_i - m_a) G} \approx 400 (T/1000)^{0.25}\, \mathrm{Gyr},
\label{mdot}
\end{equation}
which has been evaluated for CO. It would be longer still for CH$_4$.  Equation \ref{mdot} depends only weakly on temperature, and is independent of other planetary parameters. The long time scale in Equation \ref{mdot} means that selective escape of hydrogen from other gases is effectively impossible when H$_2$ atmospheres evaporate. Of course this statement does not apply to selective escape of gases with respect to condensates, which can form droplets that rain down through the wind.

\subsection{Partial Atmospheric Evaporation} \label{evap}
While we have so far assumed H/He envelopes with uniform heavy element mass fractions, this may well be too simple.  Atmospheric evaporation can expose deeper atmospheric layers which could be enhanced in metals due to a composition gradient.  Is has long been suggested that Uranus and Neptune may feature a composition gradient, with the fluid planetary ``ice'' becoming gradually more abundant with depth \cp[e.g.][]{Hubbard91}.  Such interior structures with composition gradients may suppress convection and could lead to thermal evolution models that better match Uranus's current \teff,  compared to the three-layer models in \S \ref{unsec}.  Whether or not there are well-defined layers, it is certainly true that the H/He mixing ratio decreases with depth in both planets.  It is worth exploring a small range of models for Uranus and Neptune with a composition gradient to assess how quickly $Z_{\rm env}$ may rise with depth inside a $\sim$15 \me\ planet.  We can use the methods of  \citet{Nettelmann08,Nettelmann13} to suggest possible interior structures.

Figure \ref{UN2} shows viable models (meaning all observational constraints are satisfied) of the structure of Uranus and Neptune can be calculated that have a gradient in $Z_{\rm env}$.  These use a solar $Z$ in the visible atmosphere, but a gradient in log (pressure) below that.  Instead of a sharp change in $Z$ at the interface between layers, the value of $Z$ can rise sharply as a function of depth.  If such models are close to reality, and if a planet with a structure like this were placed into a relatively close-in orbit, some amount of the outer layers would be lost, exposing a higher $Z_{\rm env}$ as time goes on.  While this is not a point that we can quantify for exoplanets at this time, it seems likely that mass loss from sub-Neptune and Neptune-class planets could lead to a potentially dramatic increase in $Z_{\rm env}$ with time.

\section{Mass-Radius at High $Z_{\rm env}$}
A natural question that arises is whether such high values of $Z$ is the envelope of LMLD planets is consistent with their measured masses and radii.  This question has really only be extensively investigated for GJ 436b \cp{Nettelmann10} and GJ 1214b \cp{Rogers10,Nettelmann11,Valencia13}.  In particular, only \ct{Nettelmann11} and \ct{Valencia13} have investigated detailed thermal evolution including different values of $Z_{\rm env}$.  Thermal evolution models are generally important when comparing to observations since the equation of state of the H-He component is quite temperature sensitive leading to significant contraction of the H-He envelope with time \cp[e.g.][]{Lopez13}.

\begin{figure} \epsscale{1.22}
\plotone{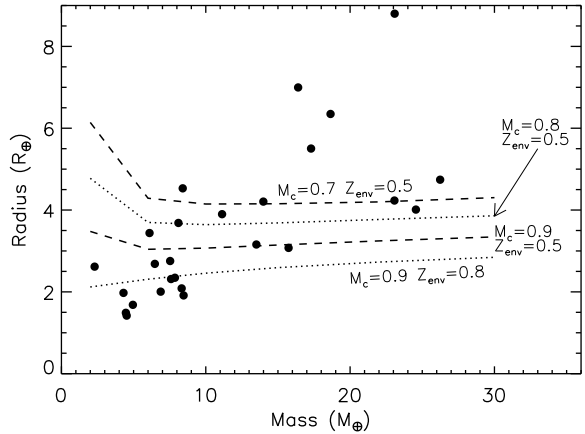}  
\caption{The mass and radius of all transiting planets with dynamically measured masses.  Model mass-radius relations are computed assuming a rock-iron core \cp{Hubbard89} that is 70-90\% of the planet's total mass ($M_{\rm c}=0.7-0.9$).  A variety of values are used for the mass of the envelope (1-$M_{\rm c}$) and its metal enrichment $Z_{\rm env}$.  Structure models use the H-He EOS of \ct{SCVH} and water equation of state of \ct{French09} for envelope.  The radiative-convective boundary is set at 100 bars in all models.  A range of  $Z_{\rm env}=0.5-0.8$ are shown.  High values of $M_{\rm c}$ and  $Z_{\rm env}$ readily match the LMLD planets in the lower left.  Generally lower values of $M_{\rm c}$ and/or $Z_{\rm env}$ are needed as planet mass increases.  Beyond $\sim$15 \me\ there is tremendous diversity in planetary properties.
\label{mr}}
\end{figure}

Calculations of thermal evolution models of all of the current LMLD planets, with a variety of $Z_{\rm env}$ values is beyond the scope of this work.  However, we can perform simpler calculations to yield plausibility to our suggestion of  metal-rich atmospheres, in particular for the lower mass planets.  In Figure \ref{mr} we plot model mass-radius curves for planets with rock-iron cores and envelopes made of H-He and water.  The core mass fraction $M_{\rm c}$ is varied from 0.9 to 0.7, while the $Z_{\rm env}$ is varied from 0.5 to 0.8, using a water EOS \cp{French09}.  We have arbitrarily set the radiative-convective boundary at 100 bars, in the absence of a lack of understanding of how interior cooling may slowed by these extremely high opacity atmospheres.  Previous work at solar metallicity for planets of similar mass suggest boundaries from several hundred bars to 1 kbar at Gyr ages \cp{Fortney07a}.

Figure \ref{mr} shows that high values of $M_{\rm c}$ and  $Z_{\rm env}$ readily match the LMLD planets in the lower left, which was already known for GJ 1214b, which can be modeled as an entirely water planet \cp[e.g.][]{Nettelmann11, Valencia13}.  As either $M_{\rm c}$ or $Z_{\rm env}$ are decreased, this leads to a larger mass fraction of H-He gas, with leads to larger radii.   For the population of known planets, generally lower values of $M_{\rm c}$ and/or $Z_{\rm env}$ are needed as planet mass increases.  This may indicate that more massive envelopes are being accreted, or that these envelopes are less metal-rich.  Beyond $\sim$15 \me\ there is tremendous diversity in planetary properties.  Kepler-30d, at 23\me\ and nearly 9 \re \cp{Sanchis12}, is perhaps 70\% H-He by mass \cp{Batygin13}.  We might naturally expect that planets like GJ 436b at 4.3 \re\ are metal-rich, while planets like Kepler-30d are relatively metal-poor.  Such a scenario is certainly observationally testable.

\section{Discussion:  Recent and Future Observations}
\subsection{GJ 436b Atmospheric Characterization}
The 22 \me\ planet GJ 436b has been observed extensively to understand better its atmosphere.  \emph{Spitzer} at occultation (secondary eclipse) has probed the dayside spectrum \cp{Stevenson10,Beaulieu11}.  A variety of observations have been made of the transmission spectrum \cp{Ballard10,Pont09,Alonso08,Caceres09,Knutson11}.  We have seen in Figure \ref{histo} that values of $\mu$ from $\sim$ 4 to 10 are quite possible.

Model interpretation of the dayside data at the time of occultation has focused on the apparent absence of CH$_4$ and abundance of CO and CO$_2$ \cp{Madhu11a,Moses12}.  In particular, preliminary calculations by \ct{Moses12} suggest that an atmosphere many hundreds of times solar (or higher) is most consistent with the derived carbon chemistry mixing ratios.  Our suggestion of high atmospheric metal mass fractions for LMLD planets certainly concurs with this work.

\begin{figure} \epsscale{1.22}
\plotone{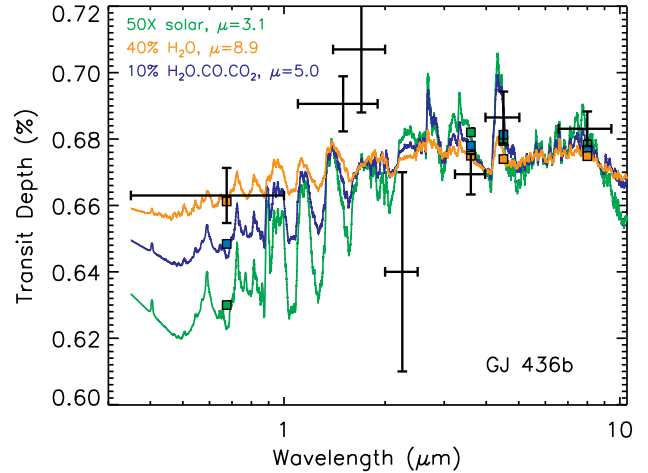}  
\caption{Model transmission spectra for three models of GJ 436b.  In green is the a 50$\times$ solar model with equilibrium chemistry.  In orange is model with a much higher $\mu$, at 50$\times$ solar model with an ad hoc inclusion of a 40\% mixing ratio of water vapor.  In blue is the same model instead with a mixing ratio of 10\% heavier molecules, evenly distributed between H$_2$O, CO, and CO$_2$.  This model better reproduces the 3.6 and 4.5 $\mu$m observations.  Model band average are shown as squares.  Higher $\mu$ models naturally better reproduce the implied relatively flat spectrum between the optical and the mid infrared.  Data from short to long wavelengths are from \ct{Ballard10}, \ct{Pont09}, \ct{Alonso08}, \ct{Caceres09}, and \ct{Knutson11}.
\label{436}}
\end{figure}

We can briefly examine model transmission spectra of the planet and compare it to observational data.  We will predominately focus on data from \emph{EPOXI} and \emph{Spitzer}, as it has the smallest error bars.  In Figure \ref{436}, shown in green, is a transmission model with a 50$\times$ solar metallicity atmosphere, and a corresponding $\mu=3.1$.  The optical transit depth is badly mis-matched by the model, while it reasonably matches the \emph{Spitzer} data.  Clearly a relatively flat spectrum would be a better match to the observations.  This immediately suggests higher $\mu$ atmospheres or those with cloud opacity.  The larger radius at 4.5 $\mu$m compared to 3.6 $\mu$m also suggests strong CO and CO$_2$ opacity \cp{Knutson11}, as was found on the dayside.

We examine two models that generally match the data better than the 50$\times$ solar model.  In orange is a model with a high $\mu$=8.9 due to a 40\% water mixing ratio, similar to a lower limit suggested by \ct{Berta12} for \gj.  This flattened transmission spectrum is a reasonable match from the optical to the mid IR.  In blue is a model with an additional 10\% mixing ratio of H$_2$O, CO, and CO$_2$, evenly distributed between the three.  The optical fit is not quite as good, but the 3.6/4.5 $\mu$m ratio is better reproduced.  We strongly concur with the findings of \ct{Knutson11} that models with strongly enhanced CO/CO$_2$, and depleted CH$_4$ (which would lead to a shallower model transit at 3.6 $\mu$m) lead to the best \emph{Spitzer} IRAC fit.  It appears that high $\mu$ atmospheres fit the available data reasonably well.

\subsection{Characterizing Atmospheric Heavy Elements in GJ 1214b with Hubble}
The atmosphere of \gj\ has generally been modeled as being of solar composition with enhancements of water vapor, or heavy inert molecules such as N$_2$ \cp{Howe12}.  The enrichment of the atmosphere of \gj, as well as our solar systems giant planets, by the accretion and dissolution of planetesimals is thought to be a generic outcome of the planet formation process.  The exact composition of such planetesimals may depend strongly on the abundances of the planet-forming disk and the location of the planets within the disk \cp{Owen99,Gautier01a,Oberg11,Madhu11b}.  Structure models of Uranus and Neptune often use an icy mixture of C-H-O-N atoms called ``synthetic Uranus" \cp{Nellis97} since more than just water is expected to dissolve into the atmosphere.

\begin{figure} \epsscale{1.22}
\plotone{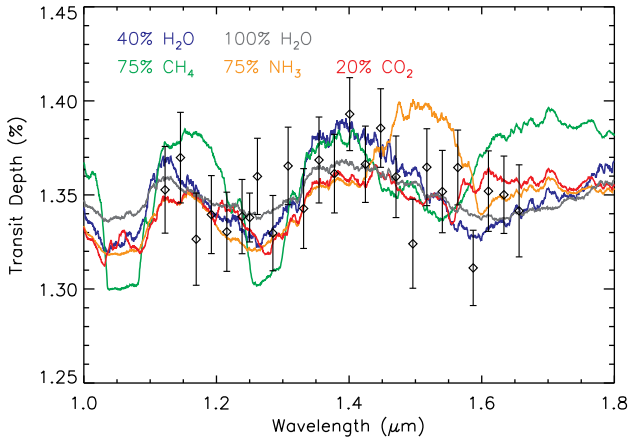}  
\caption{Several transmission spectrum models compared to the \emph{HST} data of \ct{Berta12}.  Models are of solar composition, with an increased mixing ratio of a given component, including 40\% water (blue), 20\% carbon dioxide (red), 75\% ammonia (orange), 75\% methane (green).  A 100\% water model is shown in gray.  Improved observations at these wavelengths could in principle secure detections of these molecules.  A current campaign \cp{Kreidberg13} should achieve 5$\times$ smaller error bars.
\label{berta}}
\end{figure}

As a suggestion of what might be seen with more precise \emph{HST} data for \gj, we have computed transmission spectrum models in Figure \ref{berta} where the additional, high $\mu$ ``icy'' volatile that we have mixed in are either water, ammonia, methane, or carbon dioxide.  With the current data, we do not feel it is wise to try and put limits on possible mixing ratios of molecules, and we compute these models merely as a guide to show where absorptions features are located.  However, J.~Bean is leading a 60 orbit \emph{HST} campaign that will achieve $\sim 5 \times$ higher signal-to-noise than \ct{Berta12}.  The goal of the program is to detect transmission absorption features even from a pure steam atmosphere \cp{Kreidberg13}.  Clearly with a  higher quality spectrum we may be able to detect molecular features besides those of water, which may provide insight on the details of the planet's atmosphere.  Determining the relative abundances of volatiles would be an important constraint on the planet's formation, such that heavy element enrichment should not just viewed simply as a way to decrease the atmospheric scale height.

\subsection{LMLD Planets and the James Webb Space Telescope}
The \emph{James Webb Space Telescope} (\emph{JWST}) will be the next observatory to provide transit and eclipse
spectroscopic capabilities significantly beyond \emph{HST}, and we have
simulated transit observations of our model atmospheres with its NIRSpec
\citep{Ferruit12} and MIRI instruments \citep{Wright10}. This work is similar to our previous detailed transmission spectrum modeling of \emph{JWST} observations for GJ 436b \cp{Shabram11}.

We computed the
number of photons detected using a model of the host star, transmission models of the planet, and estimates of the total efficiency
(detected electrons per incident photon) at each wavelength for the \emph{JWST}
NIRSpec $0.7 - 5$ $\mu$m and MIRI low resolution spectrograph (LRS) $R
\equiv \lambda / \delta \lambda \sim 100$ spectroscopic modes. Photon
noise was also computed, and we added it in quadrature to an assumed
observational noise floor that is consistent with current knowledge of
instrument performance.  The chosen models include three from Figure \ref{berta}:  the 100\% water model (blue), the 1$\times$solar+75\% CH$_4$ model (green), and the 1$\times$solar+20\% CO$_2$ model (red).  We also include a hazy model with 0.25$\mu$m-size soot particles (magenta) from \ct{Morley13} (their Figures 12 and 13).  
\begin{figure} \epsscale{1.22}
\plotone{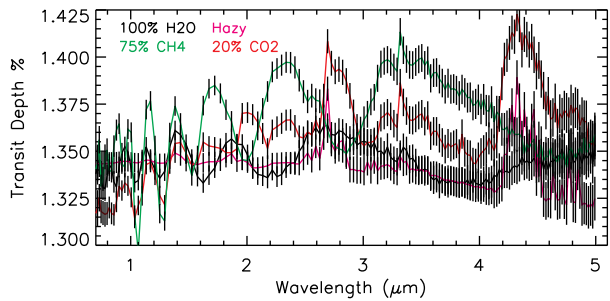}  
\setlength{\abovecaptionskip}{-210pt}
\caption{\emph{JWST} NIRSpec simulations of the transit of \gj.  The four models are 100\% water  (blue), 1$\times$solar+75\% CH$_4$ model (green), 1$\times$solar+20\% CO$_2$ model (red), all from Figure \ref{berta}, and a hazy model with 0.25$\mu$m-size soot particles (magenta) from \ct{Morley13}.  Error bars are from observing four transits.  See text for details.
\label{g1}}
\end{figure}
\begin{figure} \epsscale{1.22}
\plotone{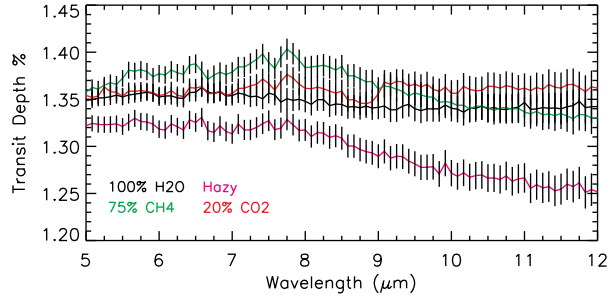}  
\setlength{\abovecaptionskip}{-210pt}
\caption{\emph{JWST} MIRI simulations of the transit of \gj.  The four models are 100\% water  (blue), 1$\times$solar+75\% CH$_4$ model (green), 1$\times$solar+20\% CO$_2$ model (red), all from Figures \ref{berta} and \ref{g1}, and a hazy model with 0.25$\mu$m-size soot particles (magenta) from \ct{Morley13}.  Error bars are from observing four transits.  See text for details.
\label{g2}}
\end{figure}

Simulated transit depths and 1 $\sigma$ error bars for observations of 4
different models are shown in Figure \ref{g1}. A total integration time of 5
hours was used in each simulation, divided into a total of 6000 s on the
star before the transit, 6000 s during transit, and 6000 s on the star
after transit. This corresponds to observing 4 transits if the
integration duty cycle is 50\%. A model of an M5V star by
\citet{Kurucz09} was used for GJ 1214, and the error bars are spaced
according to the modeled spectroscopic resolution of the 2 instrument
modes. The systematic noise floor was set to 35 parts per million (ppm)
for NIRSpec because this was recently achieved by \citet{Deming13} using
the new scanning mode of the G141 grism on \emph{HST} WFC3 (which uses a
similar HgCdTe detector). Due to the relatively coarse NIRSpec spatial sampling
below 3 $\mu$m wavelength, a similar scanning mode will have to be
implemented on \emph{JWST} to obtain a similarly low systematic noise value in
the presence of known detector intrapixel response variations and
observatory pointing errors.

However, the slitless NIRISS spectrograph \citep{Doyon12} will sample the \emph{JWST} point
spread function much more finely and will likely provide low systematic
errors at these wavelengths even without a scanning mode. A systematic
noise floor value of 50 ppm was adopted for the MIRI LRS simulations, shown in Figure \ref{g2},
the same value used by \citet{Deming09} in similar simulations. More
details of the simulation code can be found in \citet{Shabram11}.

It is interesting to note that the two flattest spectra in the WFC3 bandpass, the pure water and hazy models, only begin to significantly diverge in the mid-infrared, when the Mie scattering opacity of the small soot particles begins to wane at long wavelengths.  It certainly appears that characterizing relatively high $\mu$ atmospheres of LMLD planets with \emph{JWST} will be challenging, and we strongly encourage future simulations of exoplanet characterization with all 4 instruments.  That will give the community a better understanding of the time involved to characterize a large sample of planets.

\section{Conclusions and Outlook}
We are just now starting an era that could be termed the rise of the LMLD planets.  The ubiquity of these planets has now been shown from Kepler data \cp{Howard12,Youdin11,Fressin13,Dressing13,Morton13}.  \gj\ and GJ 436b have recently been joined by GJ 3470b \ct{Bonfils12} and HD 97658b \cp{Dragomir13} as small transiting planets around bright stars.   It is likely that soon other planets around bright stars will be found and they will be targets for detailed characterization. Formation models are now aiming to understand the frequency of such planets, their formation timescale, and how often they end up in compact orbital configurations \cp{Hansen12,Rogers11,Ikoma12,Chiang13}.  

Our experience from \gj\ shows that such planets may not give up their secrets easily.  We find that this behavior could be a generic feature of the atmospheres of these planets.  Our population synthesis models, as well as experience for Uranus and Neptune, show that large enhancements in metals may well be common for LMLD planets.  Simulations presented here show values of $Z_{\rm env}$, the metals mass fraction in the H/He envelope are commonly 0.6 to 0.9.  This will shrink atmospheric scale heights for planets in the 2-4 \re\ range.  However, there is incredible diversity within this class, far more than the diversity for planets of tens of Earth masses or higher, so some planets should have smaller $Z_{\rm env}$ values.

If LMLD planets have $Z_{\rm env}$ composition gradients, as has been suggested for Uranus, the gradual loss of the outer regions of the H/He atmosphere could lead to the excavation of atmosphere with progressively larger values of $Z_{\rm env}$.  We also find that increasing the atmospheric $\mu$ via the differential escape of hydrogen from heavy elements is unlikely due to extremely long diffusion times between H$_2$ and heavier molecules.

Extremely high mixing ratios of heavy elements will lead to vast reservoirs of material to form thick and opaque clouds \cp{Morley13}.  Condensate clouds are ubiquitous in planetary atmospheres across wide temperature ranges.  \ct{Morley13} have shown that equilibrium condensate clouds such as ZnS and KCl may be found in the atmosphere of \gj, as well as hazes derived from methane photo-destruction.  Equilibrium clouds could easily move in or out of the visible atmosphere of \gj\ analogs at hotter and cooler temperatures.  However if photochemical hazes are present, they could obscure the atmosphere across a wide range of temperatures.  A break in behavior could occur for the warmest LMLD planets, as we find that planets above \teff\ of 800-1100 K will no longer be methane dominated, but rather CO-dominated, which would end the pathway to haze formation.

In terms of future theoretical work, we make several suggestions.   The first is an exploration of the equilibrium and non-equilibrium chemistry of atmospheres strongly enriched in metals as $Z_{\rm env}$ values from 0.6 to 0.9 suggest atmospheres of 100$\times$ to 400+ $\times$ solar abundances.  Chemistry calculations will be especially important in understanding the thermal emission spectra of the planets, since dayside spectra will not suffer for high $\mu$ values diminishing the atmospheric signatures.  If planetesimal accretion includes rocky objects that do not bring hydrogen (as the ices do) even higher metallicities are possible, if these rocky planetesimals also ablate within the atmosphere.  We have shown that high atmospheric metallicities are a generic outcome of the Mordasini et al.~formation models.  It may be that across the range of LMLD planets to warm Neptune class objects like GJ 436b, that exceptionally metal rich atmospheres are very common, and should be studied in much more detail \cp[e.g.,][]{Moses12}.

Atmospheric mass loss of H/He envelopes in the extreme metal-rich atmosphere regime should also be investigated.  The cooling provided by abundant metals, in the form of molecules, atoms, and ions, should be quantified.  The high atmospheric $\mu$ together with efficient radiative cooling, could decrease atmospheric mass loss relates below those observed and modeled for H/He dominated hot Jupiter atmospheres.

With a much better understanding of planet formation and migration, one could in principle consider characterizing planetary atmospheric composition to understand better a planet's formation location and history.  Such work is now happening for gas giants \cp{Konopacky13}.  One could imagine a gradient in accreted planetesimals that depended on distance from the parent star.  In addition, planets that formed in situ inside the snow line would accrete relatively volatile-deficient planetesimals, while those that formed beyond and migrated closer in would accrete a mix of  volatile-rich and -poor planetesimals that changed with time. 

Exoplanetary science will continue to be an observation driven field.  Now that we know the LMLD planets are ubiquitous, to better characterize these planets more nearby examples are needed to provide targets for JWST and other platforms.   With the rise of specialized ground-based transit survey for M dwarfs \cp{Nutzman08,Berta12b,Giacobbe12,Sozzetti13}, and the Transiting Exoplanet Survey Satellite \cp[TESS,][]{Ricker10}, the prospects are extremely good.  We should expect to be rewarded with a rich diversity of planetary properties.
\\
\acknowledgements
JJF acknowledges support from NASA awards NNX09AC22G, NNX12AI43A, and HST-GO-12251.05.  We had many  helpful conversations with Mark Marley, Caroline Morley, and Eric Lopez.  We thank Brad Hansen and Johanna Teske for comments on an earlier draft.

\appendix

\section{Relations between mean molecular weight ($\mu$), mass fraction ($Z$), and ratios to solar abundances}
\subsection{Mean molecular weight and O:H ratio as a function of $Z$, for H-He-water mixtures}
We compute the O:H ratio and the mean molecular weight of hydrogen-helium-water mixtures with 
given mass fractions $X$ of hydrogen, $Y$ of helium, and $Z$ of heavy elements (here, water).
Further given are the molecular weights of each of these components, which we call $\mu_x$, $\mu_y$, and $\mu_z$.
The molecular weights in g mole$^{-1}$ are atomic H ($\muH=1$), helium ($\muHe=4$), and water ($\muWater=18$). 
By definition,
\begin{equation}\label{XYZ}
	X = \frac{\mu_x\,N_x}{M}\quad,\quad
	Y = \frac{\mu_y\,N_y}{M}\quad,\quad
	Z = \frac{\mu_z\,N_z}{M}\quad,
\end{equation}
where $N_x$, $N_y$, $N_z$ are the numbers of particles of these species in a given volume $V$ with total mass $M$ and total number of particles $N$, given by
\[
	M = \mu_x\,N_x + \mu_y\,N_y + \mu_z\,N_z \quad,\quad N = N_x + N_y + N_z\: . 
\]
From (\ref{XYZ}) we see the general relations
\begin{equation}\label{NxyzNxyz}
	\frac{N_y}{N_x} = \frac{Y}{X}\,\frac{\mu_x}{\mu_y}\quad,\quad
	\frac{N_z}{N_x} = \frac{Z}{X}\,\frac{\mu_x}{\mu_z}\quad\:.
\end{equation}
Turning from the general to specific relations, we next assume a mixture of hydrogen, helium, 
and water (H$_2$O) and derive the O:H ratio for two cases, case $(i)$ for atomic hydrogen, where  
$\mu_x = 1\,\muH$, case $(ii)$ for molecular hydrogen, where $\mu_x = 2\,\muH$.
In case $(i)$ we have 
\[
	N_{\rm He}=N_y\quad,\quad
	N_{\rm O}=N_z\quad,\quad
	N_{\rm H}=N_x + 2\,N_z\quad,
\]
and in case $(ii)$ we have
\[
	N_{\rm He}=N_y\quad,\quad
	N_{\rm O}=N_z\quad,\quad
	N_{\rm H}=2\,N_x + 2\,N_z\quad.
\]
So we can write down the O:H ratios as
\begin{eqnarray}
\mbox{(case $i$)}\quad{\rm O:H}  = \frac{N_{\rm O}}{N_{\rm H}} 
&=& \frac{N_z}{N_x+2\,N_z} = \frac{N_z/N_x}{1+2\,N_z/N_x} = \frac{\frac{Z}{X} (\mu_x / \mu_z)}{1+2[\frac{Z}{X} (\mu_x / \mu_z)]}  \\
\mbox{(case $ii$)}\quad{\rm O:H}  = \frac{N_{\rm O}}{N_{\rm H}} 
&=& \frac{N_z}{2\,N_x+2\,N_z} = \frac{N_z/N_x}{2+2\,N_z/N_x} = \frac{\frac{Z}{X} (\mu_x / \mu_z)}{2+2[\frac{Z}{X} (\mu_x / \mu_z)]}\:,
\end{eqnarray}
in which the $N_z/N_x$ ratios from (\ref{NxyzNxyz}) have been inserted to obtain the O:H ratio in terms of $X$ and $Z$ at the far right.  For reference, in Figure 6 in \ct{Nettelmann11}, we assumed case (ii) because in the H-He envelope of GJ1214b and LMLD atmospheres generally, hydrogen would be molecular.
The mean molecular weight reads
\begin{equation}\label{eq:barmu}
\bar{\mu} := \frac{\mu_x\,N_x + \mu_y\,N_y + \mu_z\,N_z}{N_x+N_y+N_z} 
= \frac{\mu_x + \mu_y\,(N_y/N_x) + \mu_z\,(N_z/N_x)}{1+N_y/N_x+N_z/N_x} = \frac{\mu_x + \mu_y[\frac{Y}{X}(\mu_x / \mu_y)] + \mu_z[\frac{Z}{X}(\mu_x / \mu_z)]}
{1+ [\frac{Y}{X}(\mu_x / \mu_y)] + [\frac{Z}{X}(\mu_x / \mu_z)]}\:,
\end{equation}
where again the ratios from (\ref{NxyzNxyz}) were inserted to obtain $\bar{\mu}$ as a function of $X$, $Y$, and $Z$, with $\mu_x$, $\mu_y$, $\mu_z$, chosen according to the current situation, e.g., $\mu_x=2\,\muH$ (for molecular hydrogen) and $\mu_z=\muWater$.

\subsection{$Z$ value for given composition $\rm \{N_i:H\}$ }

For a given composition $\{N_i\}$ of atomic heavy elements with respective atomic weights $\mu_i$ 
we can write down the corresponding metals mass fraction $Z$ according to its definition,
\begin{equation}
	Z = \frac{\sum_i \mu_i N_i}{{\rm \mu_{H} N_H} + {\rm \mu_{He} N_ {He}} + \sum_i \mu_i N_i}\:.
\end{equation}
Often, the composition is provided in terms particle abundance ratios $N_i$:H, thus we have
\begin{equation}
		Z = \frac{\sum_i \mu_i \mbox{$N_i$:H}}{1 + \mbox{$\mu_{\rm He}$ He:H} + \sum_i \mu_i \mbox{$N_i$:H}}\:.\label{eq:Zdef}
\end{equation}
Of course, the value of $Z$ must not change whether the abundance ratios refer to atomic hydrogen ($N_i$:H) or molecular hydrogen ($N_i$:H$_2$), where $N_i$:H$_2$ = $2 N_i$:H. We show this explicitly:
\begin{equation}
		Z = \frac{\sum_i \mu_i \mbox{$N_i$:H$_2$}}{\mu_{H_2}\mbox{H$_2$:H$_2$} + \mbox{$\mu_{\rm He}$ He:H$_2$} 
	+ \sum_i \mu_i \mbox{$N_i$:H$_2$}}
	= \frac{2\sum_i \mu_i \mbox{$N_i$:H}}{2\mu_H + 2\mbox{$\mu_{\rm He}$ He:H} + 2\sum_i \mu_i \mbox{$N_i$:H}}\:.
\end{equation}

\subsection{Mean molecular weight $\mu$ for given $Z$ value}

We can invert Eq.~(\ref{eq:Zdef}) to express the mean molecular weight of the heavy elements,
\begin{equation}
	\mu_Z := \frac{\sum_i \mu_i \mbox{$N_i$:H}}{\mbox{$\sum_i$ $N_i$:H}} 
\end{equation}
in terms of $Z$,
\begin{equation}\label{eq:muZ}
	\mu_Z = \frac{1 + \mbox{$\mu_{\rm He}$ He:H}}{\sum_i \rm \mbox{$N_i$:H}} \left(\frac{Z}{1-Z}\right)\:.
\end{equation}
When using equation (\ref{eq:muZ}) one has to make sure that the heavy element enrichment factors 
$\alpha_i=N_i$:H / $N_i$:H$_{sol}$ over the solar system abundances $N_i$:H$_{sol}$ are the same as the ones used to compute $Z$.

However, the total mean molecular weight of an atmosphere, $\mu$ does depend on whether hydrogen is in molecular or atomic form. If atomic (\emph{case i}), we have
\begin{equation}
\mu^{(H)} = \frac{1 + 4 \mbox{He:H} + \mu_Z\sum_i \mbox{$N_i$:H} }{1 + \mbox{He:H} + \sum_i \mbox{$N_i$:H}}\:;
\end{equation}
and if molecular (\emph{case ii}), we have
\begin{equation}
	\mu^{(H_2)} = \frac{\mu_{\rm H_2} \mbox{H$_2$:H$_2$} + \mu_{\rm He} \mbox{He:H$_2$} + \mu_Z\sum_i \mbox{$N_i$:H$_2$} }
		{\mbox{H$_2$:H$_2$} + \mbox{He:H$_2$} + \sum_i \mbox{$N_i$:H$_2$}}
	= \frac{2 + 4 \mbox{He:H$_2$} + \mu_Z\sum_i \mbox{$N_i$:H$_2$} }	{1 + \mbox{He:H$_2$} + \sum_i \mbox{$N_i$:H$_2$}}\:.	
\end{equation}
Both forms are consistent with Eq.~(\ref{eq:barmu}).

%



\end{document}